\documentclass[12pt]{article}

\usepackage{latexsym}

\usepackage{graphicx}

\begin{document}


\title{Electrically charged dilaton black holes in external magnetic field }

\author{
     Stoytcho S. Yazadjiev \thanks{E-mail: yazad@phys.uni-sofia.bg}\\
{\footnotesize  Department of Theoretical Physics,
                Faculty of Physics, Sofia University,}\\
{\footnotesize  5 James Bourchier Boulevard, Sofia~1164, Bulgaria
}\\ \footnotesize{and}\\
             \footnotesize  ${}^{}$ Theoretical Astrophysics, Eberhard-Karls University of T\"ubingen, T\"ubingen 72076, Germany }

\date{}

\maketitle

\begin{abstract}
In the present paper we construct a new solution to the
Einstein-Maxwell-dilaton gravity equations describing electrically
charged dilaton black holes immersed in a strong external magnetic
field and we study its properties. The black holes described by the
solution are rotating but with zero total angular momentum and
possess an ergoregion confined in a neighborhood of the horizon. Our
results also show that the external magnetic field does not affect
the black hole thermodynamics.

\end{abstract}


\sloppy

\section{Introduction}

Dilaton black holes have been extensively studied in various aspects
during the last two decades. Nevertheless,  there still remain open
problems which are not solved up to now. The present paper is
devoted to one such problem, namely the construction of exact
solutions describing electrically charged dilaton black holes in  a
strong external magnetic field and the study of their properties. In
contrast to the exact solutions describing uncharged dilaton black
holes in external magnetic field, the construction of electrically
charged black holes in external magnetic field is much more
difficult. The reason hides behind the fact that such solutions have
to be in principle stationary due to the contribution of
electromagnetic circulating momentum flux $E\wedge B$ in the
energy-momentum tensor. It is well known, however, that the
construction of rotating black holes in Einstein-Maxwell-dilaton
(EMd) gravity is extremely difficult for arbitrary dilaton coupling
parameter. In fact, the only known rotating black hole solution in
EMd gravity is the black hole with Kaluza-Klein coupling
$\alpha=\sqrt{3}$ \cite{FZ},\cite{DR}. The finding of rotating EMd
black holes with arbitrary dilaton coupling parameter is another yet
unsolved problem. In view of the described difficulties we will
consider the EMd gravity with $\alpha=\sqrt{3}$.

The study of black holes in a  magnetic field  (the so-called
magnetized black holes) has a long history starting with the
classical papers by Wald \cite{WALD} and Ernst \cite{ERN}.
Afterwards the magnetized black holes within general relativity and
other alternatives theories were studied by many authors in various
aspects ranging from pure theoretical studies to astrophysics
\cite{EW}-\cite{Yazadjiev1}.  The thermodynamics of the magnetized
black holes is of special interest. The naive expectation  was that
a thermodynamic description of these configurations would include
also the value of the background magnetic field as a further
parameter. However, more detailed studies  showed  that this was not
the case for a static electrically uncharged solutions in four
dimensions and some rotating black holes in higher dimensions -- the
external magnetic field only distorts the horizon geometry without
affecting the thermodynamics \cite{RADU}, \cite{O1},
\cite{Yazadjiev1}. The thermodynamics of electrically charged or/and
rotating magnetized black holes is not fully studied and it is not
clear whether the magnetic field affects it or not. The situation
with the thermodynamics of electrically charged and/or rotating
magnetized black holes seems very complicated in the light of the
very recent result of \cite{GMP}. Unexpectedly, it was discovered in
\cite{GMP} that the magnetized electrically charged
Reissner-Nordstr\"om and Kerr-Newman black holes are not asymptotic
to the static Melvin solution \cite{MEL1}. Even more surprising, it
was shown in \cite{GMP} that, in the general case, the ergoregion of
the magnetized Reissner-Nordstr\"om and Kerr-Newman black holes
extends all the way from the horizon to infinity. The natural
question is whether a similar behavior occurs for other electrically
charged black holes. In the present paper we give also a partial
answer to this question for some black holes in EMd gravity.

\section{Construction of the exact solution}
The field equations of the four-dimensional Einstein-Maxwell-dilaton
gravity  are  given by

\begin{eqnarray}
&&R_{\mu\nu}= 2\nabla_{\mu}\varphi \nabla_{\nu}\varphi  + 2e^{-2\alpha\varphi}  \left(F_{\mu\sigma}F_{\nu}{}^{\sigma}- \frac{g_{\mu\nu}}{4}F_{\rho\sigma}F^{\rho\sigma} \right),\\
&&\nabla_{\mu}\left(e^{-2\alpha\varphi}F^{\mu\nu}\right)=0=\nabla_{[\mu} F_{\nu\sigma]}, \\
&&\nabla_{\mu}\nabla^{\mu}\varphi=
-\frac{\alpha}{2}e^{-2\alpha\varphi}F_{\rho\sigma}F^{\rho\sigma} ,
\end{eqnarray}
where  $\nabla_{\mu}$ and  $R_{\mu\nu}$ are the  Levi-Civita
connection and the Ricci tensor with respect to the spacetime metric
$g_{\mu\nu}$. $F_{\mu\nu}$ is the Maxwell tensor and the dilaton
field is denoted by $\varphi$, with $\alpha$ being the dilaton
coupling parameter governing the coupling strength of the dilaton to
the electromagnetic field.

We shall consider spacetimes admitting a  spacelike Killing field
$\eta$ with closed orbits. The dimensional reduction of the field
equations along the Killing field was performed in \cite{Y2}. So we
present here only the basic steps skipping the details which can be
found in \cite{Y2}.

Since the Maxwell 2-form $F$ is invariant under the flow of the
Killing field there exist potentials  $\Phi$ and $\Psi$ defined by
$d\Phi=i_{\eta}F$ and $d\Psi=-e^{-2\alpha\varphi} i_{\eta}\star F$
such that

\begin{eqnarray}\label{Maxwell}
F=X^{-1}\eta \wedge d\Phi - X^{-1} e^{2\alpha\varphi} \star
\left(d\Psi \wedge \eta\right),
\end{eqnarray}
where

\begin{eqnarray}
X=g(\eta,\eta).
\end{eqnarray}
The twist $\omega$ of the Killing field $\eta$, defined by
$\omega=\star(d\eta\wedge \eta)$,  satisfies the equation

\begin{eqnarray}
d\omega= 4d\Psi \wedge d\Phi= d(2\Psi d\Phi - 2\Phi d\Psi).
\end{eqnarray}
Hence we conclude that there exists a twist potential $\chi$ such
that $\omega=d\chi + 2\Psi d\Phi - 2\Phi d\Psi$.

The projection metric $\gamma$ orthogonal to the Killing field
$\eta$, is defined by

\begin{eqnarray}
g= X^{-1}\left(\gamma + \eta\otimes \eta\right).
\end{eqnarray}
In local coordinates adapted to the Killing field, i.e.
$\eta=\partial/\partial\phi$, we have

\begin{eqnarray}
ds^2= g_{\mu\nu}dx^{\mu}dx^{\nu}= X(d\phi + W_a dx^a)^2 +
X^{-1}\gamma_{ab}dx^a dx^b.
\end{eqnarray}
The 1-form $W=W_a dx^a$ is closely related to the twist $\omega$
which can be expressed in the form

\begin{eqnarray}\label{TRF}
\omega= X i_{\eta} \star dW .
\end{eqnarray}

The dimensionally reduced EMd equations form an effective
3-dimensional gravity coupled to a nonlinear $\sigma$-model with the
following action

\begin{eqnarray}
{\cal A}= \int d^{\,3}x \sqrt{-\gamma} \left[R(\gamma) -
2\gamma^{ab}G_{AB}\partial_{a}X^A\partial_{b}X^B \right],
\end{eqnarray}
where $R(\gamma)$ is the Ricci scalar curvature with respect to the
metric $\gamma_{ab}$, $X^A=(X, \chi,\Phi,\Psi, \varphi)$ and
$G_{AB}$ can be viewed as a metric on an abstract Riemannian
manifold ${\cal N}$ with local coordinates $X^A$ and its explicit
form is given by

\begin{eqnarray}
G_{AB}dX^{A}dX^B= \frac{dX^2 + \left(d\chi + 2\Phi d\Psi - 2\Psi
d\Phi \right)^2 }{4X^2} + \frac {e^{-2\alpha\varphi}d\Phi^2
 + e^{2\alpha\varphi} d\Psi^2}{X} + d\varphi^2 .
\end{eqnarray}

What is important for the aim of the present paper is the fact that
the Riemannian space $({\cal N}, G_{AB})$ is a symmetric space for
the critical coupling $\alpha=\sqrt{3}$. In fact ${\cal N}$ is an
$SL(3, R)/O(3)$ symmetric space and therefore its metric can be
written in the form
\begin{eqnarray}
G_{AB}dX^{A}dX^B= \frac{1}{8} Tr\left(S^{-1}dS S^{-1}dS\right) ,
\end{eqnarray}
where $S$ is  a symmetric $SL(3,R)$ matrix explicitly given by
\cite{Y3}

\begin{eqnarray}
S\!=\!e^{-\frac{2}{3}\sqrt{3}\varphi}\!\!\left(
  \begin{array}{ccc}
  \! X^{-1}& -2X^{-1}\Phi & X^{-1}(2\Phi\Psi - \chi) \\
    -2X^{-1}\Phi  &  e^{2\sqrt{3} \varphi} + 4X^{-1}\Phi^2 & -2e^{2\sqrt{3}\varphi}\Psi - 2X^{-1}(2\Phi\Psi -\chi)\Phi \\
    X^{-1}(2\Phi\Psi - \chi) & -2e^{2\sqrt{3}\varphi}\Psi - 2X^{-1}(2\Phi\Psi - \chi)\Phi &  X + 4\Psi^2e^{2\sqrt{3}\varphi} + X^{-1}(2\Phi\Psi -\chi)^2 \\
  \end{array}
\!\right) .\nonumber
\end{eqnarray}

The group of symmetries $SL(3,R)$ can be used to generate new
solutions from known ones via the scheme

\begin{eqnarray}
S \to \Gamma S \Gamma^{\,T},  \;\; \gamma_{ab} \to \gamma_{ab},
\end{eqnarray}
where $\Gamma\in SL(3,R)$. In the present paper we consider seed
solutions corresponding to the  matrix

\begin{eqnarray}
S_{0}=e^{-\frac{2}{3}\sqrt{3}\varphi_{0}}\left(
        \begin{array}{ccc}
          X^{-1}_{0} & 0 & 0 \\
          0 & e^{2\sqrt{3}\varphi_{0}} & -2e^{2\sqrt{3}\varphi_{0}}\Psi_{0} \\
          0 & -2e^{2\sqrt{3}\varphi_{0}}\Psi_{0} & X_{0} + 4\Psi^2_{0}e^{2\sqrt{3}\varphi_{0}} \\
        \end{array}
      \right)
\end{eqnarray}
and transformation matrices in the form

\begin{eqnarray}
\Gamma= \left(
            \begin{array}{ccc}
              1 & B & 0 \\
              0 & 1 & 0 \\
              0 & 0 & 1 \\
            \end{array}
          \right),
\end{eqnarray}
with $B$ being an arbitrary real number. In the particular case when
the seed solution is a pure Einstein black hole solution, i.e.
$\Psi_{0}=0$ and $\varphi_{0}=0$, the above transformation gives the
solution describing  uncharged black holes in external magnetic
field \cite{Yazadjiev1}. The physical meaning of the parameter $B$
is the asymptotic strength of the external magnetic field.

The new solutions which can be generated from the seed  and the
transformation matrices under consideration, are encoded in the
matrix $S=\Gamma S_{0}\Gamma^{\,T}$ and the explicit form of their
potentials is as follows

\begin{eqnarray}
&&X=\frac{X_{0}}{\sqrt{1 + B^2 e^{2\sqrt{3}\varphi_{0}}X_{0} } } , \\
&&e^{\frac{4}{\sqrt{3}}\varphi}=
\frac{e^{\frac{4}{\sqrt{3}}\varphi_{0}}}{1 + B^2
e^{2\sqrt{3}\varphi_{0}}X_{0} } , \\
&&\Phi= -\frac{B}{2} \frac{e^{2\sqrt{3}\varphi_{0}}X_{0}}{1 + B^2
e^{2\sqrt{3}\varphi_{0}}X_{0}},  \\
&&\Psi=\Psi_{0},\\
&&\chi= -2\Psi_{0}\Phi .
\end{eqnarray}

In order to find the solution describing an electrically  charged
dilaton black hole in  external magnetic field we choose the seed
solution to be the electrically charged, static and spherically
symmetric dilaton black hole solution with $\alpha=\sqrt{3}$
\cite{GM},\cite{GHS}, namely

\begin{eqnarray}
&&ds_{0}^2 = - \frac{1- \frac{r_{+}}{r}}{\sqrt{1-\frac{r_{-}}{r} } }
dt^2 +  \frac{\sqrt{1-\frac{r_{-}}{r} }}{1- \frac{r_{+}}{r}} dr^2  +
r^2 \left(1 - \frac{r_{-}}{r}\right)^{3/2} (d\theta^2 + \sin^2\theta
d\phi^2), \\
&&e^{2\sqrt{3}\varphi_{0}}=\left(1 - \frac{r_{-}}{r}\right)^{3/2},\\
&&F_{0}= -\frac{q}{r^2} dt\wedge dr .
\end{eqnarray}

The relation between the mass and the charge of the seed solution
with the parameters $r_{+}$ and $r_{-}$ is given by the formulae
\begin{eqnarray}
m=\frac{1}{2}\left(r_{+} - \frac{r_{-}}{2}\right), \;\;\;\; q^2=
\frac{r_{+}r_{-}}{4} .
\end{eqnarray}

Using the definition of the potential $\Psi$ we find that the
potential $\Psi_{0}$ corresponding to the seed solution is given by

\begin{eqnarray}
\Psi_{0}= - q\cos\theta.
\end{eqnarray}

The only two quantities that should be found in order to obtain the
new solution describing an electrically charged dilaton black hole
in an external magnetic field, are the rotational 1-form $W$ and the
Maxwell 2-form $F$. The rotational 1-form $W$ can be found by
reversing eq.(\ref{TRF})

\begin{eqnarray}
dW= X^{-2}i_{\eta}\star \omega
\end{eqnarray}
and taking into account that for the new solution $\omega= - 4\Phi
d\Psi_{0}$. After some algebra with differential forms we obtain

\begin{eqnarray}
W= - \frac{2q B}{r} dt .
\end{eqnarray}

The Maxwell 2-form can be found in the same way by using
eq.(\ref{Maxwell}) and the result is

\begin{eqnarray}
F=-d\left[\frac{q}{r (1 + B^2 e^{2\sqrt{3}\varphi_{0}}X_{0}
)}\right]\wedge dt + d\Phi\wedge d\phi ,
\end{eqnarray}
or equivalently the gauge potential is given by

\begin{eqnarray}
A=A_{\mu}dx^{\mu} = -\frac{q}{r (1 + B^2
e^{2\sqrt{3}\varphi_{0}}X_{0} )}dt  +  \Phi d\phi .
\end{eqnarray}

Now we can present the solution in a fully explicit form:

\begin{eqnarray}
&&ds^2 = \sqrt{1 + B^2r^2 \left(1-
\frac{r_{-}}{r}\right)^3\sin^2\theta} \left[- \frac{1-
\frac{r_{+}}{r}}{\sqrt{1-\frac{r_{-}}{r} } } dt^2 +
\frac{\sqrt{1-\frac{r_{-}}{r} }}{1- \frac{r_{+}}{r}} dr^2  + r^2
\left(1 - \frac{r_{-}}{r}\right)^{3/2} d\theta^2 \right]  \nonumber \\
&& + \frac{r^2 \left(1 -
\frac{r_{-}}{r}\right)^{3/2}\sin^2\theta}{\sqrt{1 + B^2r^2 \left(1-
\frac{r_{-}}{r}\right)^3\sin^2\theta}} \left(d\phi - \frac{2qB}{r}
dt \right)^2 , \\
&&e^{2\sqrt{3}\varphi}= \left[\frac{1- \frac{r_{-}}{r}} {1 + B^2r^2
\left(1 - \frac{r_{-}}{r}\right)^{3}\sin^2\theta}\right]^{3/2}, \\
&&A_{t}= -\frac{q}{r\left[1 + B^2r^2 \left(1 -
\frac{r_{-}}{r}\right)^{3}\sin^2\theta\right]}, \\
&& A_{\phi}= -\frac{B}{2} \frac{r^2 \left(1 -
\frac{r_{-}}{r}\right)^{3}\sin^2\theta}{1 + B^2r^2 \left(1 -
\frac{r_{-}}{r}\right)^{3}\sin^2\theta} .
\end{eqnarray}

In the particular case when $q=0$ the solution reduces to the
solution describing the Schwarzschild-dilaton black hole with
$\alpha=\sqrt{3}$ in an external magnetic field.

It is interesting  to note that our 4-dimensional EMd solution
corresponds to a pure vacuum solution to the 5-dimensional Einstein
gravity which can be found by performing a Kaluza-Klein uplifting
via the equation

\begin{eqnarray}
ds^2_{5}= e^{\frac{2}{3}\sqrt{3}\varphi}  ds^{2}_{4} +
e^{-\frac{4}{3}\sqrt{3}\varphi}\left(dx_{5} +
2A_{\mu}dx^{\mu}\right)^2.
\end{eqnarray}

In completely explicit form we have

\begin{eqnarray}\label{5DSOL}
ds^2_{5}= - \left(1 - \frac{r_{+}}{r -r_{-}} \right)dt^2  +
\frac{r-r_{-}}{r-r_{+}}dr^2  + (r-r_{-})^2 d\theta^2 +
(r-r_{-})^2\sin^2\theta d\phi^2  \nonumber \\
+ \left[\frac{r}{r-r_{-}} + B^2(r-r_{-})^2\sin^2\theta
\right]dx^2_{5} - 2\frac{\sqrt{r_{+}r_{-}}}{r-r_{-}}dt dx_{5}-
2B(r-r_{-})^2\sin^2\theta d\phi dx_{5}.
\end{eqnarray}

It is also worth noting that there is one more method for generating
our EMd solution. It can be generated  via a twisted Kaluza-Klein
reduction of the 5-dimensional uplifted seed solution

\begin{eqnarray}\label{5DSeed}
&&ds^2_{5\,(seed)}= - \left(1 - \frac{r_{+}}{r -r_{-}} \right)dt^2
+ \frac{r-r_{-}}{r-r_{+}}dr^2  + (r-r_{-})^2 d\theta^2 +
(r-r_{-})^2\sin^2\theta d\phi^2  \nonumber \\
&&+ \frac{r}{r-r_{-}} dx^2_{5} -
2\frac{\sqrt{r_{+}r_{-}}}{r-r_{-}}dt dx_{5}
\end{eqnarray}
along the Killing field $V=B \frac{\partial}{\partial \phi} +
\frac{\partial}{\partial x_{5}}$. Indeed, introducing the new
coordinate $\phi_{*}=\phi - Bx_{5}$, which is  invariant under $V$,
we find that (\ref{5DSeed}) transforms to (\ref{5DSOL}) and the
further Kaluza-Klein reduction obviously gives our EMd solution.

\section{Properties of the solution}

In the present section we investigate some of the basic properties
of the solution constructed in the previous section.

First we note that our solution is free from conical singularities
and the periodicity of $\phi$ is the usual one $\Delta\phi=2\pi$. In
order to see that the parameter $B$ is indeed the asymptotic
magnetic field let us calculate  ${\vec B}^2$ on the axis of axial
symmetry. We have

\begin{eqnarray}
 {\vec B}^2|_{axis}=B^2
e^{4\sqrt{3}\varphi_{0}}=B^2 \left(1- \frac{r_{-}}{r}\right)^3 ,
\end{eqnarray}
which shows that $B$ is the asymptotic magnetic field  strength in
the limit $r\to \infty$.

There is a Killing horizon at $r=r_{+}$ where the Killing field
$K=\frac{\partial}{\partial t} + \Omega_{H}\frac{\partial}{\partial
\phi}$ becomes null. Here   $\Omega_{H}= \frac{2qB}{r_{+}}$ is the
angular velocity of the horizon. The metric induced on the horizon
cross section is

\begin{eqnarray}\label{Hmetric}
ds^2 _{H}= \sqrt{1 + B^2r^2 \left(1-
\frac{r_{-}}{r}\right)^3\sin^2\theta}\,\, r^2 \left(1 -
\frac{r_{-}}{r}\right)^{3/2} d\theta^2 + \frac{r^2 \left(1 -
\frac{r_{-}}{r}\right)^{3/2}\sin^2\theta}{\sqrt{1 + B^2r^2 \left(1-
\frac{r_{-}}{r}\right)^3\sin^2\theta}} d\phi^2 . \nonumber \\
\end{eqnarray}

By applying the Gauss-Bonnet theorem one can show that surface of
the event horizon is topologically a 2-sphere. The horizon area is

\begin{eqnarray}
{\cal A}_{H}= 4\pi r^2_{+} \left(1 -
\frac{r_{-}}{r_{+}}\right)^{3/2}
\end{eqnarray}
and evidently it coincides with the horizon area of the seed
solution. As in the uncharged case, the external magnetic field
deforms the horizon but preserves the horizon area. In fact the
geometry of the horizon cross section deviates from that of the
round 2-sphere. A simple  inspection of (\ref{Hmetric}) reveals that
the polar circumference ($\phi=const$) is greater than the
circumference about the equator ($\theta=\frac{\pi}{2}$). Therefore,
for the solution under consideration the magnetic field elongates
the black hole along the magnetic field and the black hole is
prolate in shape.

The physical electric charge is given by

\begin{eqnarray}
Q=\frac{1}{4\pi} \int_{H} e^{-2\sqrt{3}\varphi}\star F=
\frac{1}{2}\int^{\pi}_{\theta=0}d\Psi= q
\end{eqnarray}
and coincides with that of the seed solution.

The ergoregion for the solution under consideration is determined by
the region where $g(\frac{\partial}{\partial
t},\frac{\partial}{\partial t})=g_{tt}$ is positive.  The explicit
form of $g_{tt}$ is

\begin{eqnarray}
g_{tt}=  \frac{- (r-r_{+}) + (r_{+} + r_{-} -
r)(r-r_{-})^2B^2\sin^2\theta }{ r\sqrt{1- \frac{r_{-}}{r}} \sqrt{1 +
B^2r^2 \left(1- \frac{r_{-}}{r}\right)^3\sin^2\theta}}
\end{eqnarray}

It is easy to see that very close to the horizon we have $g_{tt}>0$
for $\sin\theta\ne 0$ and $g_{tt}=0$ for $\sin\theta=0$ and
$r=r_{+}$. Also, it is not difficult  to see that for   $r\ge r_{+}
+ r_{-}$ it holds that $g_{tt}<0$. Therefore, we conclude that there
exists an ergoregion confined in a compact neighborhood of the
horizon, in contrast with the magnetized Reissner-Nordstr\"om
solution for which the ergorgeon extents to infinity \cite{GMP}. The
boundary of the ergoregion, i.e. the ergosurface, is defined by
$g_{tt}=0$ which reduces to a cubic equation in $r$, namely
$r-r_{+}=(r_{+} + r_{-} - r)(r-r_{-})^2 B^2\sin^2\theta$. Solving
this cubic equation for $r$ we can find the equation $r(\theta)$ of
the ergosurface. The explicit form of $r(\theta)$ is  too cumbersome
to be presented here. What is important is that the ergosurface is
qualitatively the same as the  ergosurface of the Kerr solution.

The surface gravity $\kappa$ associated with the Killing field $K$
can be calculated via the well-known formula
\begin{eqnarray}
\kappa^2= - \frac{g(d\lambda,d\lambda)}{4\lambda}\,,
\end{eqnarray}
where $\lambda=g(K,K)$. The direct computation gives the following
result

\begin{eqnarray}
\kappa= \frac{1}{2r_{+}\sqrt{1-\frac{r_{-}}{r_{+}}}}\,,
\end{eqnarray}
which is just the surface gravity of the seed solution. Therefore,
the surface gravity is not affected by the external magnetic field
for the solution under consideration.

The inspection of the electric field shows that ${\vec E}^2 \to 0$
for $r \to \infty$. The same holds for the norm of the twist of the
Killing vectors $\eta=\frac{\partial}{\partial \phi}$ and
$\xi=\frac{\partial}{\partial t}$. As a whole, our solution is
asymptotic to the dilaton-Melvin solution with $\alpha=\sqrt{3}$.
This can be easily seen from its explicit form.

\section{Thermodynamics}
The study of the thermodynamics of the black holes in external
magnetic fields is difficult because of the  asymptotic structure.
Since the spacetime is not asymptotically flat a substraction
procedure is needed to obtain finite quantities from integrals
divergent at infinity. The natural choice  for the substraction
background in our case is the dilaton-Melvin background. To
calculate the mass we use the quasilocal formalism
\cite{Yazadjiev1}. Here we give for completeness a very brief
description of the quasilocal formalism.

The spacetime metric is  decomposed into the form
\begin{equation}
ds^2 = - N^2dt^2 + h_{ij}(dx^i + N^{i}dt)(dx^j + N^{j}dt),
\end{equation}
with $N$ and $N^{i}$ being the lapse function and  the shift vector.
The decomposition means that the spacetime is foliated by spacelike
surfaces $\Sigma_{t}$ of metric $h_{\mu\nu} = g_{\mu\nu} +
u_{\mu}u_{\nu}$, labeled by a time coordinate $t$ with a unit normal
vector $u^{\mu} = - N\delta^{\mu}_{0}$. The spacetime boundary
consists of the initial surface $\Sigma_{i}$ ($t=t_{i}$), the final
surface $\Sigma_{f}$ ($t=t_{f}$) and a timelike surface ${\cal B}$
to which the vector $u^{\mu}$ is tangent. The surface ${\cal B}$ is
foliated by $2$-dimensional surfaces $S^{r}_{t}$, with metric
$\sigma_{\mu\nu}= h_{\mu\nu} - n_{\mu}n_{\nu}$, which are the
intersections of $\Sigma_{t}$ and ${\cal B}$. The unit spacelike
outward normal to $S^{r}_{t}$, $n_{\mu}$, is orthogonal to
$u^{\mu}$.

In order to have well-defined variational principle we must consider
the extended  EMd action with the corresponding boundary terms
added:

\begin{eqnarray}\label{SEMDA}
S= {1\over 16\pi} \int d^4x \sqrt{-g}\left(R -
2g^{\mu\nu}\partial_{\mu}\varphi \partial_{\nu}\varphi  -
e^{-2\alpha\varphi}F^{\mu\nu}F_{\mu\nu} \right) \nonumber \\ +
{1\over 8\pi}\int_{\Sigma_{i}}^{\Sigma_{f}} K\sqrt{h}d^3x -{1\over
8\pi}\int_{{\cal B}}\Theta \sqrt{\sigma}d^{2}x.
\end{eqnarray}
Here $K$ is the trace of the extrinsic curvature $K^{\mu\nu}$ of
$\Sigma_{t_{i,f}}$ and $\Theta$ is the trace of the extrinsic
curvature $\Theta^{\mu\nu}$ of ${\cal B}$, given by

\begin{eqnarray}
K_{\mu\nu} &=& - {1\over 2N}\left( {\partial h_{\mu\nu}\over \partial t} - 2D_{(\mu}N_{\nu)} \right) ,\\
\Theta_{\mu\nu} &=& - h^{\alpha}_{\mu} \nabla_{\alpha} n_{\nu},
\end{eqnarray}
where $\nabla_{\mu}$ and $D_{\nu}$ are the covariant derivatives
with respect to the metric $g_{\mu\nu}$ and $h_{\mu\nu}$,
respectively.

The quasilocal energy $M$ and the angular momentum $J_{i}$ are given
by

\begin{eqnarray}
M = {1\over 8\pi} \int_{S^{r}_{t}} \sqrt{\sigma} \left[N(k-k_{0}) +
{n_{\mu}p^{\mu\nu}N_{\nu}\over \sqrt{h}} \right] d^{D-2}x \nonumber
\\ +
{1\over 4\pi}\int_{S^{r}_{t}} A_{0} \left({\hat \Pi}^{j} - {\hat \Pi}_{0}^{j} \right)n_{j}d^{D-2}x ,\\
J_{i} = - {1\over 8\pi} \int_{S^{r}_{t}} {n_{\mu}p^{\mu}_{i}\over
\sqrt{h} } \sqrt{\sigma}d^{D-2}x - {1\over 4\pi} \int_{S^{r}_{t}}
A_{i}{\hat \Pi}^{j}n_{j}d^{D-2}x .
\end{eqnarray}

Here $k= - \sigma^{\mu\nu}D_{\nu}n_{\mu}$ is the trace of the
extrinsic curvature of $S^{r}_{t}$ embedded in $\Sigma_{t}$. The
momentum variable $p^{ij}$ conjugated to $h_{ij}$ is given by

\begin{equation}
p^{ij} = \sqrt{h}\left(h^{ij}K - K^{ij} \right).
\end{equation}

The quantity ${\hat \Pi}^{j}$ is defined by

\begin{equation}
{\hat \Pi}^{j} = - {\sqrt{\sigma}\over \sqrt{h}}
\sqrt{-g}e^{-2\alpha\varphi}F^{0j}.
\end{equation}

The quantities with the subscript ``0'' are those associated with
the background.

After  long calculations, for the quasilocal energy of the black
hole we find
\begin{eqnarray}
M=m=\frac{1}{2}(r_{+} - \frac{1}{2}r_{-})
\end{eqnarray}
which is evidently independent from the external magnetic field and
coincides with the mass of the seed solution. In the same way, for
the angular momentum we obtain
\begin{eqnarray}
J=0.
\end{eqnarray}
Even more, the angular momentum $J_{S^{r}_{t}}$ associated with
every surface $S^{r}_{t}$ is zero. So we have a rotating black hole
(i.e. $\Omega_{H}\ne 0$) while the  total angular momentum is zero.
This, at first sight strange result, means that the gravitational
contribution to the angular momentum  is exactly compensated by the
opposite in sign angular momentum of the electromagnetic field.

Furthermore, the following Smarr-like relation is satisfied
\begin{eqnarray}
M= \frac{1}{4\pi}\kappa {\cal A}_{H} + \Xi_{H} Q ,
\end{eqnarray}
where the potential $\Xi_{H}$ is in fact the corotating electric
potential evaluated on the horizon and given by

\begin{eqnarray}
\Xi_{H}= - K^{\mu}A_{\mu}|_{H}= - (A_{t} +
\Omega_{H}A_{\phi})|_{H}=\frac{Q}{r_{+}} .
\end{eqnarray}

On the basis of the results obtained  so far we can conclude that
the external magnetic field does not affect the thermodynamics of
the charged dilaton black holes.

\section{Discussion}

In the present paper we constructed a new solution to the EMd
gravity equations describing charged dilaton black holes in external
magnetic field for  dilaton coupling parameter $\alpha=\sqrt{3}$.
The basis properties of the solution and its thermodynamics were
studied. The black holes described by the solution are rotating but
with zero total angular momentum and possess an ergoregion confined
in a neighborhood of the horizon. Our results also show that the
external magnetic field does not affect the black hole
thermodynamics.

The natural generalization of this work is to consider electrically
charged and rotating dilaton black holes immersed in an external
magnetic field and to study their thermodynamics. Especially, it is
interesting whether the thermodynamics of the rotating solutions
depends nontrivially on the external magnetic field. Other
interesting questions are the asymptotic structure at infinity and
the compactness or non-compactness of the ergoregion. These problems
are currently under investigation and the results will be presented
elsewhere.

\vspace{1.5ex}
\begin{flushleft}
\large\bf Acknowledgments
\end{flushleft}
The author is grateful to the Research Group Linkage Programme of
the Alexander von Humboldt Foundation for the support of this
research and the Institut f\"ur Theoretische Astrophysik T\"ubingen
for its kind hospitality. He also acknowledges partial support from
the Bulgarian National Science Fund under Grant DMU-03/6.

\end{document}